\documentclass[a4paper,12pt]{article}
\usepackage[cp1251]{inputenc}
\usepackage[english]{babel}
\usepackage{amsfonts}
\usepackage{amssymb}
\usepackage[hyperindex,breaklinks]{hyperref}

\title{Time-like singularities in General Relativity}
\author{Serge~L.~Parnovsky \thanks{e-mail: par@observ.univ.kiev.ua} \\
Astronomical Observatory of \\ Taras Shevchenko National University of Kyiv \\
Observatorna str., 3a, Kyiv, 04053, Ukraine}
\date{}

\begin{document}
\maketitle
\begin{abstract}
We review the properties of naked time-like singularities in the General 
Relativity and quantum effects in their vicinity. We demonstrate that only 
line-like singularities can be formed by a collapse and are the only candidates 
to break the Cosmic Censorship Principle.
\end{abstract}

This is a brief review of the properties of naked time-like singularities in 
the General Relativity, partially based on our results obtained since late 
1970s. First of all let us note that a singularity in this paper means a region 
of a space-time where some curvature invariants diverge, but not in a 
$\delta$-function fashion. Thus we drop out all the pathological, directional, 
and conical singularities. Some directional singularities will appear later, 
but only as a by-product. If a hypersurface, which lies infinitely close to the 
singularity, is space-like than we deal with the space-like singularity, e.g. 
the cosmological or the Schwarzschild singularity. If such a hypersurface is 
time-like, the singularity is also time-like. There are two possibilities 
regarding their appearance. If there is an even number of horizons around a 
time-like singularity, then it is covered by these horizons. To an external 
observer this object looks like a black hole. An example of such an object is 
the Reisner-Nordstr\"om black hole with two horizons. An observer cannot see the 
singularity from the outside. If there are no horizons at all, this time-like 
singularity is called a naked singularity. 

There are a lot of exact solutions of the Einstein equations with such 
singularities. But there is a problem with this type of singularity: a distant 
observer can see it. It could inject radiation, matter and information. This 
prevents setting a Cauchy problem for our space-time if it contains at least 
one naked singularity without the knowledge of the boundary condition on it. To 
discard this problem, Penrose proposed the so-called Cosmic Censorship 
Principle \cite{ref:Penrose}. It states that all singularities produced by a 
collapse must be inside the horizons. Theoretically, it tolerates naked 
singularities produced by the Big Bang, but during the inflation they must fly 
away from the visible part of the Universe. So the practical conclusion is that 
there are no naked singularities in our Universe inside a cosmological horizon. 
Nevertheless the Cosmic Censorship Principle is just a hypothesis. In any case, 
currently we have no actual possibility to prove or disprove it by studying the 
process of the collapse. For this reason we proposed to study the properties of 
the naked singularities in the General Relativity in order to get some 
conclusions in this regard. Some results of this study are set out below.

Let us give a brief note on the cosmological constant. Nowadays we believe in 
the existence of the cosmological constant or dark energy, which acts similarly 
to it. However, its influence on a metric vanishes in the vicinity of time-like 
singularities. So, one can neglect the cosmological constant when studying 
naked singularities.

We begin from typology and hierarchy of naked singularities and give some 
examples. The simplest example with the metric depending only on one spatial 
coordinate $x$ is the spatial Kasner solution \cite{ref:Kasner}
\begin{equation}\label{eqn:Kasner}
ds^2=-dx^2+x^{2p_1}dt^2-x^{2p_2}dy^2-x^{2p_3}dz^2
\end{equation}
with one negative and two positive Kasner indices $p_i$ satisfying the 
conditions
\begin{equation}
p_1+p_2+p_3=1,
\end{equation}
\begin{equation}
p_1^2+p_2^2+p_3^2=1.
\end{equation}
In the paper \cite{ref:ParKhal} it was identified as the metric around the 
infinitely long straight thread with the constant linear mass density. Let us 
start from the Weil metric describing static axially symmetric space-times:
\begin{equation}
ds^2 = e^\nu dt^2 - \rho^2 e^{-\nu} d\phi^2 - 
e^{\gamma-\nu} \left(d\rho^2 + dz^2\right),
\end{equation}
\begin{equation}
\frac{1}{\rho}\frac{\partial}{\partial\rho} 
\left(\rho\frac{\partial\nu}{\partial\rho} + \frac{\partial^2\nu}{\partial 
z^2}\right) = 0,
\end{equation}
\begin{equation}
\frac{\partial\gamma}{\partial z} = 
\rho\frac{\partial\nu}{\partial\rho}\frac{\partial\nu}{\partial z},
\end{equation}
\begin{equation}
\frac{\partial\gamma}{\partial z} = 
\frac{\rho}{2}\left[\left(\frac{\partial\nu}{\partial\rho}\right)^2 - 
\left(\frac{\partial\nu}{\mathstrut\partial z}\right)^2\right].
\end{equation}
Here a function $\nu(\rho,z)$ is a harmonic axially symmetric function. In the 
conditional flat space with the cylindrical coordinate set $\rho$, $\phi$, $z$ 
it describes the Newtonian potential of some axially symmetric mass 
distribution. Assuming the source to be a thread $\rho=0$ with constant linear 
mass density $\mu$, the Kasner metric (\ref{eqn:Kasner}) can be derived after 
the transformation $x=\rho^{\mu^2-\mu+1}$ with
\begin{equation}
p_1=\frac{\mu}{\mu^2-\mu+1},
\end{equation}
\begin{equation}
p_2=\frac{1-\mu}{\mu^2-\mu+1},
\end{equation}
\begin{equation}
p_3=\frac{\mu^2-\mu}{\mu^2-\mu+1}.
\end{equation}
However at $\mu>1$ we get some problems. To resolve them let us consider the 
case of the source in the form of a finite thread $\rho=0$ with constant linear 
mass density $\mu$ and the length $L$. Using the oblate spheroid coordinate set 
$v$, $u$, $\phi$, one can easily obtain the function $\nu=2\mu \ln(\tanh(v/2))$ 
and, introducing it into the Weil metric, get the Zipoy-Voorhees metric 
\cite{ref:Zipoy, ref:Voorhees}
\begin{equation}\label{eqn:zv}
\begin{array}{l}
ds^2=\tanh^{2\mu}\frac{v}{2}dt^2\\
- \frac{L^2}{4}\tanh^{-2\mu}\frac{v}{2}\sinh^2v 
\left[\left(1+\frac{\cos^2u}{\sinh^2v}\right)^{1-\mu^2}(dv^2+du^2) + 
\cos^2ud\phi^2\right].
\end{array}
\end{equation}
The naked singularity corresponds to $v=0$. But what is the type of this 
singularity? Is it line-like, point-like or something else? This problem was 
investigated in paper \cite{ref:Par85} by using diagrams, describing the 
simplest properties of a space-time. It was shown that the case $\mu<0$ 
corresponds to a point-like singularity with negative mass, the case $0<\mu<1$ 
corresponds to a line-like singularity with positive mass. In the most complex 
case $\mu>1$ we deal with the new type of singularity, which was named 
paradox-like in paper \cite{ref:Par88}, with positive mass. If $\mu\ge2$ there 
are two additional directional singularities on its ``ends'' $v=0$, 
$u=\pm\pi/2$. In this case the space-time (\ref{eqn:zv}) has three different 
spatial infinities.

The metric (\ref{eqn:zv}) was generalized for the case $\mu\ne const$, but all 
the generalized solutions are approximate near a singularity. If $\mu$ depends 
on $z$, we get Weil singularities \cite{ref:Par85}, if it depends on $z$ and on 
$t$ -- ``simple line sources'' \cite{ref:Israel}, if it depends on $z$, on $t$, 
and on $\phi$ -- the generalized spatial Kasner metric \cite{ref:Par80}
\begin{equation}\label{eqn:gk}
ds^2=-dx^2+(x^{2p_1}l_il_k-x^{2p_2}m_im_k-x^{2p_3}n_in_k)dx^idx^k.
\end{equation}
All these solutions are approximate at $x\to0$ or $r\to0$. Their properties 
were analyzed in \cite{ref:Par85}. It was shown that these solutions describe 
either a point-like singularity with negative mass or a line-like one with 
positive mass or a paradox-like one. However, point-like singularities repulse 
collapsing matter and cannot be formed by a collapse. Paradox-like 
singularities must have a linear density exceeding the critical value and also 
cannot be formed by a collapse. Therefore, only line-like singularities could 
be considered as a candidate to break the Cosmic Censorship Principle.

But all these solutions are not general enough. The general solution of 
Einstein equations near a time-like singularity was found and analyzed in the 
paper \cite{ref:Par80}. It is an oscillating solution (naturally, an 
approximate one) very similar to the well-known Belinsky-Khalatnikov-Lifshitz 
(BKL) solution near space-like singularities \cite{ref:BKL}. In order to get a 
general solution near an arbitrary singularity it has to be matched with the 
BKL solution. This was done in the paper \cite{ref:Par90}.

An influence of non-gravitational fields was analyzed in the papers 
\cite{ref:Par88, ref:ParGayd} and some other ones. Only scalar fields can 
``kill'' a general oscillatory metric. In this case the generalized spatial 
Kasner metric (\ref{eqn:gk}) with all positive indices is the most general 
solution near a naked singularity. In the case of a real collapse we have to 
take into account quantum effects. If a classical collapse leads to the 
formation of a naked singularity, it could cause a strong radiation due to the 
quantum pair production and changing of vacuum polarization. Its backreaction 
could slow the collapse in such a way, that it forms a black hole instead of a 
naked singularity. Thus we need to calculate the mass loss due to quantum 
radiation during a formation of the naked singularity. A simple model with a 
massive shell, shrinking up to the Planck length was used. The main conclusion 
is that the mass loss is very small at the formation of a linear singularity 
\cite{ref:Par79}, but very large at the formation of a 
Reisner-Nordstr\"om singularity with $Q>M$ \cite{ref:Par81a}.

Also already formed naked singularities could be ``dressed up'' due to quantum 
radiation and its backreaction. For example, a naked Kerr singularity with 
$a>M$ acts in this way \cite{ref:Par81b}. We have the Kerr metric with the mass 
$M$ and the angular momentum $J=Ma$. If $a>M$ it describes a naked singularity. 
Quantum particles production leads to a decrease of the angular momentum $J$ 
and its mass $M$. Their ratio $a$ decreases faster than $M$, so a naked Kerr 
singularity can turn into a rotating black hole. An estimation of the time of 
``dressing up'' in the Planck units is
\begin{equation}
T_\textrm{dress} \sim 
50M\alpha^9\exp(2\alpha^2)\ln\left(\frac{M^2}{\alpha}\right),
\end{equation}
\begin{equation}
\alpha = \frac{a}{M} > 1.
\end{equation}
During the time since the Big Bang, a singularity with a solar mass can ``dress 
up'' if $1+\varepsilon < \alpha < 4$, where $\varepsilon \sim 10^{-20}$.

Coming back to the Cosmic Censorship Principle we can make some brief 
conclusions. Some types of naked singularities cannot be formed by a collapse. 
Point-like singularities repulse collapsing matter. Paradox-like singularities 
must have a linear density exceeding the critical value. General oscillatory 
singularities must have a strong influence of the quantum effects. Although we 
cannot estimate this influence, it is possible that the formation of such 
singularities is also forbidden due to these effects. However, line-like 
singularities have no such problems. So, in order to prove or disprove The 
Cosmic Censorship Principle one has to study a collapse with the formation of a 
line-like naked singularity.

\end{document}